\documentclass[acmsmall,nonacm]{acmart}

\AtBeginDocument{%
  \providecommand\BibTeX{{%
    \normalfont B\kern-0.5em{\scshape i\kern-0.25em b}\kern-0.8em\TeX}}}

\begin{document}

\title{NFTrig: Using Blockchain Technologies for Math Education}

\author{Jordan Thompson}
\email{jordanthompson18@augustana.edu}
\affiliation{%
  \institution{Augustana College}
  \city{Rock Island}
  \country{USA}}
  
\author{Ryan Benac}
\email{ryanbenac18@augustana.edu}
\affiliation{%
\institution{Augustana College}
\city{Rock Island}
\country{USA}}

\author{Kidus Olana}
\email{kidusolana18@augustana.edu}
\affiliation{%
  \institution{Augustana College}
  \city{Rock Island}
  \country{USA}}
  
\author{Talha Hassan}
\email{talhahassan18@augustana.edu}
\affiliation{%
  \institution{Augustana College}
  \city{Rock Island}
  \country{USA}}

\author{Andrew Sward}
\email{andrewsward@augustana.edu}
\affiliation{%
  \institution{Augustana College}
  \city{Rock Island}
  \country{USA}}

\author{Tauheed Khan Mohd}
\email{tauheedkhanmohd@augustana.edu}
\affiliation{%
  \institution{Augustana College}
  \city{Rock Island}
  \country{USA}}


\begin{abstract}
NFTrig is a web-based application created for use as an educational tool to teach trigonometry and block chain technology. Creation of the application includes front and back end development as well as integration with other outside sources including MetaMask and OpenSea. The primary development languages include HTML, CSS (Bootstrap 5), and JavaScript as well as Solidity for smart contract creation. The application itself is hosted on Moralis utilizing their Web3 API. This technical report describes how the application was created, what the application requires, and smart contract design with security considerations in mind. The NFTrig application has underwent significant testing and validation prior to and after deployment. Future suggestions and recommendations for further development, maintenance, and use in other fields for education are also described. 
\end{abstract}


\maketitle

\section{Introduction}
The purpose of this report is to describe the technical details involved in the development of the NFTrig application. This includes both the front end website design, the back end smart contract, and NFT creation. It will mainly focus on the technical details of the project outlining software requirements, design through programming languages, client and server side interactions, and validation testing. This allows the reader to undertake further development, fixes, or maintenance of the software, as this forms part of the documentation for the
software.

The NFTrig project is based around the creation of a web-based game application that allows interaction of NFTs (non-fungible token) with trigonometric function designs. NFts are digital assets, for example a picture, that has a  unique identification and can generally be freely traded with cryptocurrency \cite{wang2021non}.  Through this application, users are able to purchase digital artwork of many different trigonometric functions and combine them using mathematical operations. Current supported operations include multiplication and division of the trigonometry functions, and the output of each operation is a new NFT card that would be the result of an operation. The old cards will then be removed from the user's possession and burned using the smart contact. For example, if a user combined the two cards Sin(x) and Cos(x) using multiplication, they would lose their two old cards and receive the new card Tan(x). Further, the NFT cards are assigned one of the following rarity levels: common, uncommon, rare, and legendary. The probability of each of these levels is defined later in this report.

The application also allows a user to connect to MetaMask, a digital wallet capable of storing a user's cryptocurrency and NFTs as well as a way to connect to block chain. The NFTrig application can also display the NFTs owned by the user and allow them to connect to OpenSea to sell the NFTrig cards on a public marketplace. The application is hosted on Moralis employing their Web3 API. Technical languages used in this project, which will be discussed in detail throughout this paper, include front end web development languages HTML, CSS (specifically Bootstrap5), and JavaScript as well as the back end smart contract development language Solidity.

In order to attract users, this application also allows a user to answer trivia questions and gain experience points. These points can then be used to unlock new sets of NFT cards or upgrade existing cards in a user's wallet. This game-like design should appeal to a younger audience and encourage them to answer trigonometry or math based questions. This will have an incredible educational benefit for the user because they will be both learning and playing a game simultaneously.

\section{Motivation}

The purpose of this application is as an educational tool for students who are attempting to understand the ways that trigonometric functions interact with each other. As opposed to just graphing these functions by hand, students will be able to generate new NFTs by combining whatever trigonometric functions they already own. In fact, using technology is shown to influence and better educational processes by increasing interaction between those in the classroom \cite{elmessiry2021nft}. Technology is becoming increasingly prevalent in every sphere of daily life, so the use of technology in a classroom setting is not only logical, but it increases the educational benefit of students \cite{raja2018impact}. However, as the technology continues to evolve, "the gap between traditional course material taught to students in B.S./M.S. programs at universities and the cutting edge of technology used in industry is widening at an unprecedented rate" \cite{rao2019developing}. By creating this project, it will give students the opportunity to gain experience with block chain, and hopefully be a starting place for narrowing that ever growing gap. After much research, it is likely that this proposed application is the first of its kind that utilizes NFTs to teach mathematical concepts. 

Aside from user benefit of this application, there is also an intellectual merit in the block chain and education fields. Best described by Carmen Holotescu, "As education becomes more open, diversified, democratised, and decentralised, the block chain technology is taken in consideration by researchers, teachers and institutions, to maintain reputation, trust in certification, and proof of learning" \cite{holotescu2018understanding}. Further, development of this project continues research on NFT and block chain technologies. This application can also serve as the boilerplate basis for other NFT-based educational tools and resources. Research for this project provides opportunities for training computer science students on how to use NFTs in general, but more specifically in educational contexts. 

NFTrig was developed by computer science students as a final senior inquiry project at Augustana College. In conjunction and with funding by the Department of Mathematics and Computer Science, this project employs a variety of software development skills and techniques that further the research and understanding of the block chain and web development field.

\section{Related Work}

Block chain technology has enabled the formation of decentralized distributed records of digital data which does not require any third party to moderate any transactions \cite{10.1007/978-3-030-14802-7_17}. The decentralized nature of block chain also renders it easy for use in a ranging variety of applications in several fields such as healthcare \cite{holbl2018systematic}, internet of things \cite{conoscenti2016blockchain}, gaming \cite{attaran2019blockchain}, banking \cite{cocco2017banking}, and education (explored in greater detail in subsection 3.1). Non Fungible token (NFTs) are a relatively new phenomena within the field of block chain based technologies, but its application in aforementioned fields are already being studied. Specifically within the healthcare context, NFT's are solving long term issues such as storing patients' private data more safely as well as maintaining better records while giving better autonomy and privacy to both patients and healthcare providers \cite{doi:10.1126/science.abm2004}. The application of NFTs in education is still an understudied area. These next related work sections explore the broader use of block chain based technologies for educational purposes, gamification, and overall collaborative learning.

\subsection{Block chain Based Technologies for Educational Purposes}

There has been extensive work concerning how block chain based technologies are enabling better ownership and sharing of personal records for students and supporting collaborative learning environments. Yumna et al. conducted a systematic literature review of the use of block chain technologies in educational sector \cite{yumna2019use}. They also propose several uses of existing block chain based technologies in educational sector that leverage the decentralized and traceable consensus making mechanisms of block chain. Researchers have examined the use of block chain to allow students to maintain educational records such as transcripts, credentials, diplomas, and learning activities \cite{grech2017blockchain, chen2018exploring, skiba2017potential}. Similarly, research has also explored learning management systems design based on block chain based technology. The technology can potentially verify a students records as well as enable the design of an automatic decentralized enrollment system which does not require moderation from school staff \cite{skiba2017potential}.

Another elegant use of block chain in the field of education is the ability to support life-long learning applications. The educational sector is becoming more diverse with a variety of different types of classrooms and learning modalities. E-learning has also allowed students to acquire licences and accreditation online. Therefore, it is imperative to maintain the learning journeys of students over time to understand the different types of learning that they have been engaging in and improving on over time. The traceable nature of block chain based technologies (defined as one of the salient features in the aforementioned systematic review by \cite{yumna2019use}) enables all of these applications.

The decentralized nature of block chains coupled with the consensus making algorithms also makes it suitable for collaborative environments. Prior research has looked at how block chain based technologies can enable better developmental experiences in the realm of business \cite{garcia2020using} but there is very minimal work on its application within the field of education application\cite{bucea2021blockchain}. 

\subsection{Applications in Education Application and Collaborative Learning}

Although preliminary in nature, limited prior work has explored the utilization of NFTs for designing various different independent learning environments for students. There are some proposed commercial systems that have analogous functioning to some of the systems described in the prior section. For example, commercial systems are looking at leveraging NFTs to award ``Pass" status to students for different courses \footnote{A teacher at Pepperdine University using NFTs to award course completion certifications to students: https://upcea.edu/tech-trends-in-higher-ed-metaverse-nft-and-dao/}. NFTs enjoy a key advantage over conventional block chain technologies as they are typically designed using the more secure Ethereum block chain enabling an even more secure record and identity management. Researchers have shown that there is promise in using NFTs as academic tokens to represent student transcripts and other records as well that can be more easily verified \cite{elmessiry2021nft}. However, there is still a dearth of academic literature in this field.

Student incentivization is heavily advocated in pedagogical literature \cite{gass2019using}. NFTs make it easier to tie incentivization to learning outcomes as they can be automatically acquired by students at any time upon completion of learning outcomes. This gives NFTs based certifications an advantage over the more traditional learning settings where students have to strongly adhere to semester timelines. Elmessiry et al. has looked at designing an incentive mechanism that can be used by teachers and students to achieve better learning outcomes in an effective and cost-efficient manner \cite{elmessiry2021nft}. They also concluded there was better engagement outcomes for students. On several metrics of usability, the students reported more than 80\% preference for buying, using, and collecting NFTs. Such independent learning methods were particularly more useful during the COVID-19 pandemic to accommodate the need of remote independent learning options. Architecturally, this project takes inspiration from \cite{elmessiry2021nft}, and applies it to a more narrower, focused domain of learning mathematical operations in this study. Further, these NFTs are also easier to share on social media \cite{kapoor2022tweetboost}. Therefore, it also allows students to more readily share their accomplishments. 

\subsection{Gamification to Support Mathematical Learning}
Since the proposed application teaches mathematical and trigonometric formulas to students, the literature on use of gamification to support mathematical learning should be better described. Gamification, in combination with incentivization explained in the previous section, will allow for the success of this application. Gaming settings have traditionally been used to teach simple mathematical operations to students. More recently, researchers have also proposed systems that teach  advanced concepts to students including College Algebra \cite{FAGHIHI2014182}. These learning environments make it easier for students to relate the learning concepts with more daily life phenomena. While gamification itself cannot guarantee better learning outcomes, it can improve students' interest and performance by encouraging them to engage with the content for a longer duration of time \cite{JAGUST2018444}. The simpler, more systematic, and operational nature of mathematics as a subject also makes it easier for incorporation in gaming environments because final answers are usually short and numerical as opposed to long and descriptive answer that might be found in social or natural sciences. Trigonometry especially can easily be broken down into a series of operations and steps which simulates a similar environment found in other online games where users play to find different ``rewards" and ``collectables". Despite all these benefits there are some limitations of gamification as well. For example, it is hard to know how a student arrived a solution and give feedback \cite{8363455}. Not being able to solve trigonometric equations can also lead to frustration and impeded learning experience. Foresight into the project's future looks to mitigate these concerns by fostering better communication between different game players and providing links to useful learning resources in the application. Prior research has extensively explored the use of gamification in different mathematical fields. This application is likely the first to extend the use of NFTs and block chain to aid in teaching trigonometric equations.

Research shows that technology, specifically games are shown to be excellent educational tools. In fact, "one of the most successful
positive reinforcement mechanisms [in education] known is gamification" \cite{elmessiry2021nft}. This includes taking a topic transforming it into a game with positive reinforcement. This leverages educational benefits in students and encourages them to continue playing the game to learn. Nftrig has future plans to add a game function which will allow the user to answer trigonometry trivia and math questions. This will aid in both their learning and the continued use of the NFTrig application. Further, the ability to combine owned NFTs with math functions also aids in the education of trigonometry for the student.

\section{Experimental Setup}
\subsection{Software Development Requirements}
The NFTrig application employs a variety of software development requirements that cover the range of the project. From front end web development to back end smart contract creation and NFT storage, this section describes the requirements and software used to complete the project. 

\subsubsection{Compiling IDE}
The smart contracts created for NFTrig are hosted on Remix. Remix is an an open source online compiler IDE that can be used to test and deploy smart contracts \cite{amir2020remix}. The platform can be accessed by any browser, and it allows the developer to write and deploy smart contracts on an actual or test server simultaneously. The current deployment is on a test server. In order to  test and debug the smart contract, Visual Studio Code is used. Visual Studio was found to be the best code editor because a developer can easily upload most file types, and edit them \cite{johnson2012professional}. For NFTrig, it was used to develop front end HTML and CSS files, as well as back end solidity contract editing. The required installed plugins for Visual Studio (VS) include Solidity and Block chain development. \cite{9565983} These allowed for simple, straightforward development of code.

\subsubsection{Moralis}
Moralis SDK is the primary back end platform for the project. The platform allows connection of the front end web application to the smart contract. \cite{9202574} The Moralis platform uses a combination of server management and a JavaScript SDK to allow for maximum interaction and simplicity. A developer can do many tasks through this including authentication of users, getting necessary user data, and connecting with MetaMask in a non-complicated and simply coded process. The only expectation is that a developer will need to have programming knowledge in JavaScript as well as a familiarity with Moralis and MetaMask, experience querying a database, and some knowledge of Web3 development to ensure maximum results and efficiency. Moralis also has the ability to easily connect to MetaMask.

\subsubsection{MetaMask}
MetaMask is the digital wallet required for participation in the NFTrig game application. It allows the collection of purchases from the user, and it can be installed as an extension on a browser for increased ease of use \cite{9354310}. MetaMask stores all NFTs owned by the user, and in connection with the NFTrig application, can view and upgrade or modify existing NFTs at a users discretion. Connection to the browser extension is required for the application to access anything owned by a user \cite{lee2019using}. Because MetaMask is easily integrated into Moralis, and thus NFTrig, there is little a user needs to do to create a connection aside from installing the MetaMask extension, and clicking connect.

\subsubsection{Front End Design}
Front end design was accomplished primarily through Visual Studio. The Live Server extension was installed which allows each developer to "host" their developed website using a native web application. Doing so allowed simplified testing and front end development. Instead of creating CSS files from scratch, the NFTrig interface heavily employs Bootstrap5, which simplifies the process of modifying the content layout and design of buttons and other content \cite{lepage1992exploring}. Moralis and Bootstrap5 each have extensive documentation to understand and support front end web development. These tools have been utilized to a near maximum extent.

\subsubsection{Web Hosting Platform}
 The initial testing of NFTrig, as previously explained, was hosted on a local live server through Visual Studio. After initial development, the project was moved to a web server hosted by Augustana College so that initial testing could begin. It is currently unclear how the site will ultimately be hosted. One option for hosting the web application is directly through Google \cite{STANDING2002151}. This would allow the website to be named something easily searchable and accessible. A second option would be to host directly through Moralis, but a limitation of this would be a more diluted website naming convention along with a more confusion process of uploading and modifying website content. Currently, the NFTrig application will remain on the local Augustana College Server.

\section{Software Design}
This section covers all of the decisions necessary to understand the development of NFTrig, as well as the technical implementation of each technology used in the design process.

\subsection{Software Architecture}
The architecture of this project follows the model-server design architecture \cite{oluwatosin2014client}. Using this model, the clients send transactions and requests to a proxy smart contract stored on the block chain which then makes the appropriate calls to the logic smart contract which is also stored on the block chain. This style of architecture is required for this project because the smart contracts must be stored on the server-side chain in order to be functional. The use of proxy contracts also allows our smart contracts to be fully upgradeable with any future updates that may need to be implemented.

\subsection{Choice of Programming Language}
This section examines and explains the benefit of each chosen language employed in NFTrig. Front end languages include HTML and CSS and the back end includes Solidity and JavaScript. Each has been chosen because they were found to be the best option for development.

\subsubsection{Solidity}
Solidity is the programming language of choice when it comes to coding smart contracts. Solidity is "similar to JavaScript and yet has some features of object-oriented languages such as Java and C++" \cite{mohanty2018basic}. This is a leading language for the development of smart contracts and use on block chain technologies. This project utilizes the solidity library openzeppelin in order to create a solid foundation for the smart contracts. Hardhat and Node JS are then used for the testing and deployment of the smart contracts to the Polygon blockchain.

\subsubsection{JavaScript}
In the NFTrig application, JavaScript (JS) is primarily used in the front end application. The primary purpose of this language is generally to create dynamic and interactive web content \cite{haverbeke2018eloquent}. For the client, JS was used in the navigation bar to allow for clickable links and resizing of the navigation bar in smaller screens. This language was also used to give buttons functionality ranging from logging in to MetaMask to purchasing NFTrig cards. Further, JS was used to test the logic of the front-end combination page until the smart contract was applied. Aside from augmenting HTML and CSS application pages, JavaScript is also used in this project to connect the back end smart contract with the from end web application. This application was also developed using Next JS and deployed via an application known as vercel.

\subsubsection{HTML and CSS}
Web development of the user interface was primarily completed using HTML and CSS (Bootstrap5). These languages are equally popular and necessary to develop the web pages \cite{ghulam2021konseptual}. Instead of creating all CSS requirements from scratch, Bootstrap5 was utilized to allow for cleaner design across web pages and better alignment of web page elements. Bootstrap5 also simplifies the need to explicitly code buttons and other interactive items. 

\subsection{Security Considerations}
\begin{figure}[!htb]
    \centering
    \includegraphics[height=6cm]{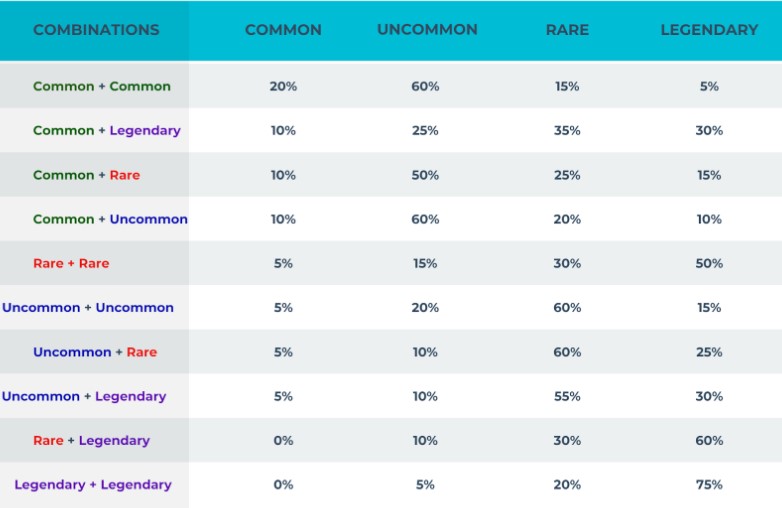}
    \caption{Probabilities of outcomes depending on rarity of selected cards}
    \label{probabilities}
\end{figure}

Throughout this project, there have been several security considerations discovered that threatened the safety and use of the application. One such discovered issue was initially, there was no code written to block a user from looking at another users token. Further, before minting a new NFT card, the smart contracts check to ensure that the card does not already exist, the cards used for combining are owned by the user, and that the newly minted card follows the correct probabilities of outcomes shows in \ref{probabilities}. These probabilities are coded into the smart contract.  

\subsection{Smart Contract Design}
The smart contract for this project is broken up into two separate contracts. The first of which is the NFTrig logic contract which contains the logic for purchasing packs of cards as well as the logic for how cards will interact with each other. The second contract is the marketplace contract which will allow users to trade their own NFTs with other users through the website. Within the NFTrig contract, there are functions for multiplying and dividing cards, purchasing randomized packs of cards, and tracking the details of each individual token as transactions are made. The marketplace contract contains information about sale history as well as the functionality to post new sales and purchase items for sale. Both of these contracts were deployed as upgradeable contracts so they can have updates implemented in the future.

\subsection{NFT Storage and Naming Conventions} 
All NFT images are stored on the server with the HTML, CSS, and JS files. The naming convention for each image references what image it is in four numbers. The first number is the power of sin, the second is the power of cos, the third is the rarity or color of the card (0-3 is green, blue, purple, and red respectively), and the final number is the text variant (0-3). These files were named accordingly to better determine the output if cards were combined using a mathematical function. For example, a sin card might have the naming convention: 1023.jpg. 10 defines it is a sin card, 2 defines it is rarity purple, and 3 defines it is text variant 3. The purpose of naming the files in this way is so that the front end can easily determine which image corresponds to a particular NFT by simply looking at the four features of each token which match the four numbers in the file name.

\subsection{Client Design} 
The NFTrig application interface was designed using HTML and CSS. The primary use of CSS was often replaced by Bootstrap5. Bootstrap 5, a library for CSS, allows for easier scaling and alignment of objects in the HTML file, and thus the computer screen \cite{krause2020introduction}. Documentation on the Bootstrap5 has utilized to a full extent. Each section examines the layout and use of each application page.

\subsubsection{NFTrig Home}
The interface is designed to allow a user to access the marketplace, their individual current collections, and their profile. The navigational bar contains links to the client-side facing pages: NFTrigHome, MyCards, CombineCards, Marketplace, and Game. We used a total of three colors to enable good contrast and make it easier for our users to view complex graphs and formula without a cluttered background \footnote{Background-color:\#333, Color: \#f2f2f2,       }. The JavaScript elements declared are reusable across multiple screens. They support functions and interactions such as a user hovering over a cell or clicking a cell and providing both feedback and error handling to the user. The navigation bar is also, the top bar changes color to indicate the tab that the user is on.
 

\subsubsection{Combination}
\begin{figure}[!htb]
    \centering
    \includegraphics[height=6cm]{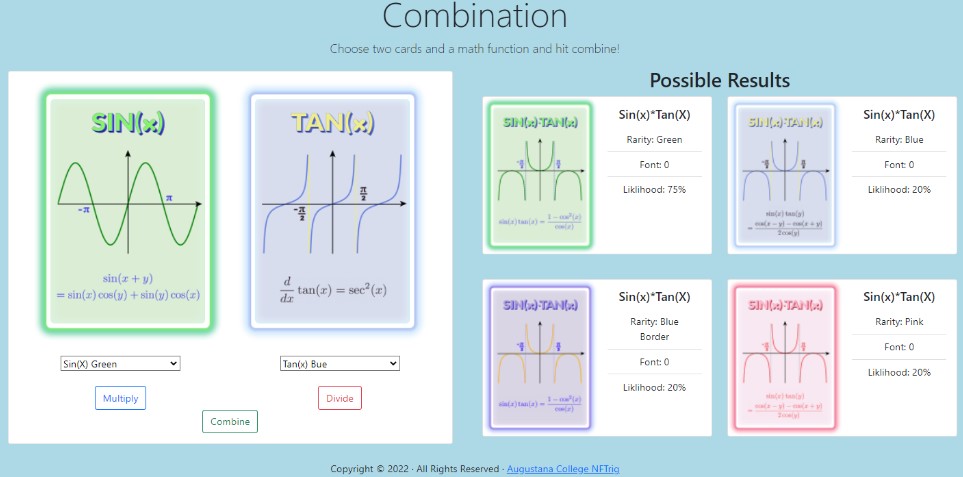}
    \caption{Interface where users will combine NFTrigs}
    \label{combination}
\end{figure}

The main purpose of the combination page is for users to choose cards that they currently own, and see options for combining them using either multiplication or division. 
Figure \ref{combination} displays the layout of the screen where user selected cards are shown on the left, and potential results are shown on the right.

The page utilizes Bootstrap5 capabilities to format effectively to different screen sizes and resolutions. It connects with a back end script to the smart contract. This provides functionality to the buttons and easy generation of possible NFT results. Below shows the probabilities of generated NFT outcomes based on the selected input cards. 

\subsubsection{Marketplace and MyCards}
\begin{figure}[!htb]
    \centering
    \includegraphics[height=6cm]{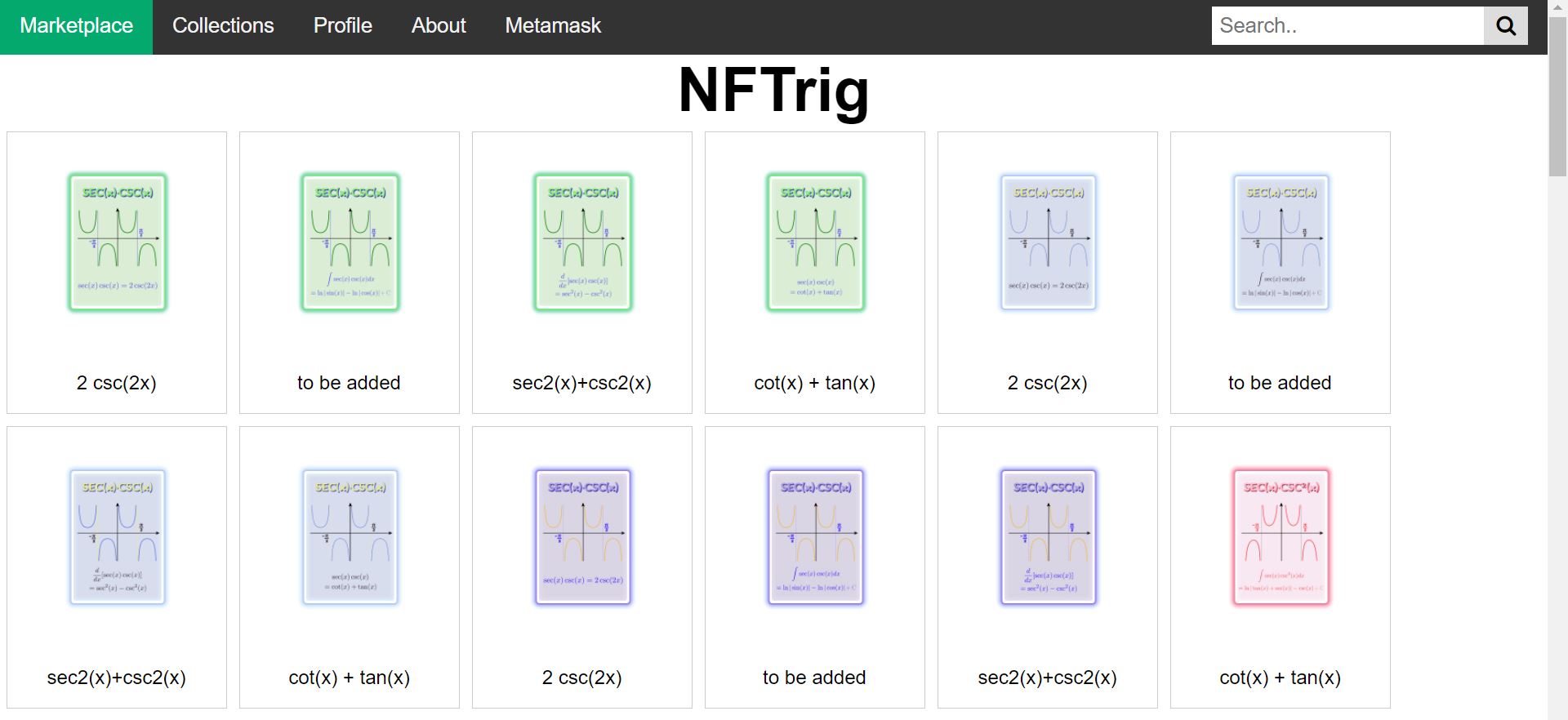}
    \caption{Interface displaying NFTrig Marketplace}
    \label{mycards}
\end{figure}

Marketplace and MyCards are similar pages, as they connect to a data source and display NFTs. The Marketplace tab shows all NFT cards available for purchase both from other users who own NFTs and cards owned by the NFTrig project. MyCards however specifically shows all cards owned by a user. The layout for each generates all necessary NFT images and information about the rarity. The rarity is signified by the color and the text option of the card. Figure \ref{mycards} shows the actual layout displayed on the page.  

\subsubsection{Quality attributes of client-side interface and code}
In order to have an application of quality, consistency, and accuracy, the project followed the following guidelines:
\begin{enumerate}
    \item The code is written in a manner that components and layouts can be rearranged to support any structural changes in the front end.
    \item The code has consistent style and format, such as the padding used in individual NFTrig elements and the purchase page's color.
    \item The code contains comments and is well indented for easy maintenance and understanding.
    \item Consistent colors and feedback systems are provided so the system is easy to learn for users.
    \item Page-level styling was avoided when possible to keep design consistent.
    \item Thorough testing was completed for basic accessibility features.
    
\end{enumerate}

\subsubsection{Testing the Client Design}
Basic unit testing of different elements was initially conducted to ensure easy navigation between front end pages. In order to ensure that testing would cover most application uses, three user cases were devised: a user browsing NFTrigs, a user making a purchase, and a user combining NFTrigs. All assumptions and expected actions expected from the system were listed and analyzed through testing. Further, testing through some edge cases were also pursued. Currently, the application works as intended, however future plans involve rigorous testing with JavaScript code and external APIs (if any are devised). This will ensure a fully functional, secure, and usable application that can also be used as a boiler plate project for other educational blockchain technologies.

\subsubsection{Future Work: Game}
Future work for this project will include the ability for users to play a trivia and trigonometric equation game. This allows a user to gain experience points that they can then use to purchase new NFTs. This eliminates the need to always need cryptocurrency to purchase individual or group NFT cards. Although there is not currently an interface for this page written in HTML, functionality exists for the trivia game itself. The files are currently stored on the server, but they are disabled and there is no navigable way to get there through the application.

\section{Methods}
Most methods for completing this project have been thoroughly explained in the sections above. However, the final intended version of this project will be hosted in a different location than it resides currently. The initial portion of this project had the front end website hosted on a local Augustana College server and the back end smart contract hosted on the Polygon test net. This allowed initial testing and validation that the smart contract operated as expected, as well as give time and opportunity to discover security vulnerabilities. The future of this project will be hosted on a decentralized web application online so that users can access it and begin to interact with the smart contract. Further, a redesign of the website user interface is likely. This will require transition from BootStrap5 to NextJS which allows cards to be generated, displayed, and interactable through a version of JavaScript.

\section{Results}
This project successfully allowed the exploration and creation of applying NFT and block chain technology to math education. Although preliminary in use and nature, this project allows for initial project creation as a boiler plate project. The smart contract is currently deployed on the Polygon testnet and can be interacted with using test Matic. Each web page has functionality to display the user's owned NFTs as well as the NFTs they have put for sale on the marketplace. Using NextJS will also allow the Combination page to have functionality and smart contract use. It is also worth noting that the created web page is not required to interact with the NFTrig smart contracts.

\section{Recommendations for Future Work}
The goal for this project was a working Beta demo that shows application functionality, and correct smart contract execution. There are many other features planned for the continued work of this project. The first, as earlier explained, is a game option which challenges the user with trigonometry trivia and math problems. Answering these questions successfully will increase the experience points of a user. The user can then use these experience points to purchase individual or packs of NFTrig cards, or they can be used to combine cards. 

\bibliographystyle{ACM-Reference-Format}
\bibliography{references}

\end{document}